\documentstyle{mn}

\newif\ifAMStwofonts
\def\gsim{\,\lower2truept\hbox{${>\atop\hbox{\raise4truept\hbox{$\sim$}}}$}\,}
\def\lsim{\,\lower2truept\hbox{${<\atop\hbox{\raise4truept\hbox{$\sim$}}}$}\,}
\def\exp{{\rm exp}}
\def\erfc{{\rm erfc}}
\def\ln{{\rm ln}}

\def\espon{{\rm exp}}
\def\der{{\rm d}}
\def\rhoav{\overline\rho}
\def\deltac{\delta_{\rm c}}
\def\lamdac{\lambda_{3{\rm c}}(z)}

\def\mass{{\rm M}}
\def\mstar{{\rm M_*}}
\def\msol{\mass_\odot}

\def\sig2{\sigma^2(\mass)}
\def\mson{\mass_{\rm o}}
\def\msonk{\mass_{{\rm o}k}}
\def\nson{\rm N_o}
\def\avmson{\overline{\mass}_{\rm o}}
\def\zson{z_{\rm o}}

\def\deltacs{\delta_{\rm co}}
\def\sigs{\sigma^2_0}
\def\mprog{\mass_{\rm p}}
\def\nprog{\rm N_p}
\def\avmprog{\overline{\mass}_{\rm p}}
\def\zprog{z_{\rm p}}

\def\deltacp{\delta_{\rm cp}}
\def\sigp{\sigma^2_{\rm p}}
\def\tform{t_{\rm f}}
\def\zform{z_{\rm f}}
\def\avzform{\overline{z}_{\rm f}}

\def\crange{[\rm{M_{i}}, \rm{M_{s}}]}

\def\rangenum{[10^i,\,10^{i+1}[}
\def\minf{\rm{M_{i}}}
\def\msup{\rm{M_{s}}}
\def\nvalm{\mathcal{N}}
\def\xih{\xi_{hh}}
\def\xim{\xi_m}
\def\b1eff{\overline{b}_1}
\def\b2eff{\overline{b}_2}
\def\bJeff{\overline{b}_J}
\ifoldfss
  \ifCUPmtlplainloaded \else
    \NewTextAlphabet{textbfit} {cmbxti10} {}
    \NewTextAlphabet{textbfss} {cmssbx10} {}
    \NewMathAlphabet{mathbfit} {cmbxti10} {} % for math mode
    \NewMathAlphabet{mathbfss} {cmssbx10} {} %  "   "    "
  \fi
  \ifAMStwofonts
    \ifCUPmtlplainloaded \else
      \NewSymbolFont{upmath} {eurm10}
      \NewSymbolFont{AMSa} {msam10}
      \NewMathSymbol{\upi}     {0}{upmath}{19}
      \NewMathSymbol{\umu}     {0}{upmath}{16}
      \NewMathSymbol{\upartial}{0}{upmath}{40}
      \NewMathSymbol{\leqslant}{3}{AMSa}{36}
      \NewMathSymbol{\geqslant}{3}{AMSa}{3E}

       \let\le=\leqslant
       
    \fi
  \fi
\fi % End of OFSS

\ifnfssone
  \newmathalphabet{\mathit}
  \addtoversion{normal}{\mathit}{cmr}{m}{it}
  \addtoversion{bold}{\mathit}{cmr}{bx}{it}
  \newmathalphabet{\mathbfit} % math mode version of \textbfit{..}
  \addtoversion{normal}{\mathbfit}{cmr}{bx}{it}
  \addtoversion{bold}{\mathbfit}{cmr}{bx}{it}
  \newmathalphabet{\mathbfss} % math mode version of \textbfss{..}
  \addtoversion{normal}{\mathbfss}{cmss}{bx}{n}
  \addtoversion{bold}{\mathbfss}{cmss}{bx}{n}
  \ifAMStwofonts
    \ifCUPmtlplainloaded \else
      \UseAMStwoboldmath
      \makeatletter
      \new@mathgroup\upmath@group
      \define@mathgroup\mv@normal\upmath@group{eur}{m}{n}
      \define@mathgroup\mv@bold\upmath@group{eur}{b}{n}
      \edef\UPM{\hexnumber\upmath@group}
      \new@mathgroup\amsa@group
      \define@mathgroup\mv@normal\amsa@group{msa}{m}{n}
      \define@mathgroup\mv@bold\amsa@group{msa}{m}{n}
      \edef\AMSa{\hexnumber\amsa@group}
      \makeatother
      \mathchardef\upi="0\UPM19
      \mathchardef\umu="0\UPM16
      \mathchardef\upartial="0\UPM40
      \mathchardef\leqslant="3\AMSa36
      \mathchardef\geqslant="3\AMSa3E

       \let\le=\leqslant

    \fi
  \fi
\fi % End of NFSS release 1

\ifnfsstwo
  \DeclareMathAlphabet{\mathbfit}{OT1}{cmr}{bx}{it}
  \SetMathAlphabet\mathbfit{bold}{OT1}{cmr}{bx}{it}
  \DeclareMathAlphabet{\mathbfss}{OT1}{cmss}{bx}{n}
  \SetMathAlphabet\mathbfss{bold}{OT1}{cmss}{bx}{n}
  \ifAMStwofonts
    \ifCUPmtlplainloaded \else
      \DeclareSymbolFont{UPM}{U}{eur}{m}{n}
      \SetSymbolFont{UPM}{bold}{U}{eur}{b}{n}
      \DeclareSymbolFont{AMSa}{U}{msa}{m}{n}
      \DeclareMathSymbol{\upi}{0}{UPM}{"19}
      \DeclareMathSymbol{\umu}{0}{UPM}{"16}
      \DeclareMathSymbol{\upartial}{0}{UPM}{"40}
      \DeclareMathSymbol{\leqslant}{3}{AMSa}{"36}
      \DeclareMathSymbol{\geqslant}{3}{AMSa}{"3E}

       \let\le=\leqslant

    \fi
  \fi
\fi % End of NFSS release 2

\ifCUPmtlplainloaded \else
  \ifAMStwofonts \else % If no AMS fonts
    \def\upi{\pi}
    \def\umu{\mu}
    \def\upartial{\partial}
  \fi
\fi

\title{Merging History Trees for Dark Matter Halos: Tests of the Merging Cell
Model in a CDM Cosmology}
\author[B. Lanzoni]
       {B. Lanzoni$^1$, G. A. Mamon$^{1,2}$, B. Guiderdoni$^1$\\
$^1$ Institut d'Astrophysique de Paris, (CNRS UPR 0341), Paris, France\\ 
$^2$ DAEC (CNRS UMR 8631), Observatoire de Paris, Meudon, France}
\date{Accepted.  
      Received ;
      in original form }

\pagerange{\pageref{firstpage}--\pageref{lastpage}}
\pubyear{1994}

\begin{document}

\maketitle

\label{firstpage}

\begin{abstract}
The merging history of dark matter halos is computed with the Merging Cell
Model proposed by Rodrigues \& Thomas (1996).
While originally discussed in the case of scale--free power spectra, it is
developed and tested here in the framework of the cold dark matter cosmology. 
%The method is based upon an actual realisation of the initial density field,
%and halos 
%are identified following the spherical model for the collapse of overdensity 
%perturbations. 
%A system of overlapping grids and some specific merging criteria are used to 
%follow the merging history of halos. 
%Structures with a large variety of masses and shapes are obtained, and also
%information about their spatial correlations are retained.
%Unconditioned and conditional halo mass functions, as well as formation
%time and largest progenitor redshift distributions have been computed and
%compared to the avalaible analytic formulae.
The halo mass function, the mass distribution of
progenitors and child halos, as well as the probability distribution of
formation times,  
%time and largest progenitor redshift distributions 
have been computed and compared to the available analytic predictions.
The halo auto--correlation function has also been obtained 
(a first for a semi--analytic merging tree), and 
%it has been
tested against analytic formulae. 
An overall good agreement is found between results of the model, and
the predictions derived from the Press \& Schechter theory and its extensions. 
More severe discrepancies appear when formulae that better describe N-body
simulations are used for comparison. 
%Apart some weakness, 
In many instances, the model can be a useful tool for
following the hierarchical growth of structures. 
In particular, it is suitable for addressing the
issue of the formation and evolution of galaxy clusters, as well as
the population of Lyman--break galaxies at high redshift, and 
their clustering properties.
%, can be studied through this method. 

%These mostly concern objects at the two ends of the mass function, with 
%the model systematically underproducing both low and high mass halos. 
%While the problem for small structures is probably inherent to the method, 
%that for massive halos is possibly due to the condition adopted for
%identifying bound objects in the Lagrangian space.
\end{abstract}


\section{Introduction}
In the basic picture of the formation of cosmic structures, the Universe is
dominated by a 
dark matter (DM) component, and small perturbations in the initial density
field grow in amplitude proportionally to a linear growth
factor, until they approach unity.
Then, non-linear effects dominate their
evolution, and the regions stop expanding with the Universe, collapse, and
virialise, thus forming DM halos. 
In hierarchical scenarios, like the cold dark matter (CDM) model,
small--scale inhomogeneities collapse first, and 
then aggregate via merging to generate larger structures.
Since galaxies form by the collapse and cooling of baryonic gas within DM
halos, and their history is greatly influenced by that of their
surrounding halos (e.g., Lemson \& Kauffmann 1999), it is important to
understand how these objects form and evolve with time. 

The most realistic way for following the history of DM halos
is by means of N--body simulations, but they require huge amounts of RAM
memory and are computationally expensive. Therefore, they are often limited
to a modest dynamic range, and to few different cosmological scenarios. 

The simplest alternative approach is to consider only the linear regime of
growth of density fluctuations, and describe the non--linear evolution and 
collapse by means of the spherical `Top-Hat' model (Gunn \& Gott 1972).  
In this formalism, the formation of a DM halo of mass M at redshift $z$ is
described by identifying in the initial density field smoothed on a scale M,
and {\it linearly} extrapolated to redshift $z$, a region
having overdensity equal to a given threshold value.
% $\deltac^{\rm ext}$.
% which shows that the collapse of an isolated spherical perturbation 
% occurs at the time when its density contrast reaches a critical value
% $\deltac$, corresponding to a value $\deltac^{\rm ext}$ of the {\it linearly
% extrapolated} density field. 
% It is therefore possible to describe the formation of DM halos in identifying
% virialised objects with those regions in the linearly extrapolated  density
% field having overdensity in excess with respect to $\deltac^{\rm ext}$.
Starting with Gaussian initial conditions, Press \& Schechter (1974,
hereafter PS) interpreted the probability of finding such regions as the
number density of halos of mass M, that formed at redshift $z$ (see Sec. 3.1). 
The PS mass function has been extensively tested against N--body simulations,
and has been found to be in reasonably good agreement with numerical results
(e.g., Efstathiou et al. 1988; Carlberg \& Couchman 1989; Lacey \& Cole 1994;
Gelb \& Bertschinger 1994). 
However, systematic deviations both at small and high masses have been 
recognised, with the PS formula predicting too many low mass halos, and
underestimating the number of massive objects (e.g., Jain \& Bertschinger
1994; Gross et al. 1998; Somerville et al. 1998; Tormen 1998; Lee \&
Shandarin 1999; Sheth \& Tormen 1999; and references therein). 
A better agreement with numerical results is obtained when ellipsoidal rather
than spherical collapse models are considered (Monaco 1995, 1997a, b; Bond \&
Myers 1996; Audit, Teyssier \& Alimi 1997, 1998; Lee \& Shandarin 1998, 1999,
hereafter LS98 and LS99; Sheth \& Tormen 1999; Sheth, Mo, \& Tormen 1999; and
references therein).  

Extensions of the PS theory (Bower 1991; Bond et al. 1991) 
follow the redshift evolution of the halo population as a whole, by deriving
the conditional probability of 
finding progenitors of mass $\mprog$ at redshift $\zprog$, given their
child halos of mass $\mson$ at $\zson$, and vice versa (see Sec. 3.2).
By means of the extended Press \& Schechter (EPS, hereafter) theory, the
distribution of halo formation and survival times, as well as their merger
rate, can also be obtained (Lacey \& Cole 1993, 1994, hereafter LC93
and LC94; see also Sec. 3.3).  
The few comparisons between these analytic predictions and numerical results
reveal a general good agreement, even if discrepancies similar to those
of the mass function have been pointed
out (LC94; Somerville et al. 1998; Tormen 1998).
 
Still based on the EPS theory, analytic predictions for halo bias in the
Lagrangian space of initial conditions have been obtained (Mo \& White 1996,
MW96 hereafter; Catelan et al. 1998, CLMP; Porciani et al. 1998; Sheth \&
Lemson 1999b; Sheth, \& Tormen 1999; Sheth, Mo, \& Tormen 1999; and references
therein). 
The halo auto--correlation function $\xih(r)$ is then the product of the halo
bias with the correlation function of the underlying matter. 
The predicted $\xih$ is in good agreement with that
in N--body simulations for massive objects, but its amplitude is too
large for low--mass halos (Porciani, Catelan \& Lacey 1999, PCL
hereafter; Jing 1999; Sheth \& Tormen 1999).  
However, Jing (1999) propose an empirical fitting formula (see Sec. 3.5) 
that provides a good description of halo clustering on the whole range of
masses (see also Sheth \& Tormen 1999; Sheth, Mo, \& Tormen 1999). 

While the PS and EPS formalisms describe the mean statistical properties of
the population as a whole, several models of the individual
merger history of DM halos have been proposed (Cole \&
Kaiser 1988; Kauffmann \& White 1993, KW93 hereafter; Rodrigues \& 
Thomas 1996, RT96 hereafter; Somerville \& Kolatt 1999; Sheth \& Lemson
1999a).
Each model presents some advantages and some drawbacks with respect to the
others. For example, the ``block model'' of Cole \& Kaiser (1988) partly
retains the spatial information, but it is affected by the discretisation of
both halo masses (in powers of two), and positions. 
The KW93 merging tree presents a more continuous spectrum of masses, 
but a grid of collapse redshifts is imposed, and the relative positions of
halos are unknown. Moreover, it reproduces exactly the mean progenitor mass
distribution, but mass conservation is enforced only approximately, while the
opposite holds in the model of Somerville \& Kolatt (1999). 

In this paper, we focus on the ``Merging Cell Model'' (MCM) \footnote{as
first coined by Nagashima \& Gouda (1997).} proposed by Rodrigues \& Thomas 
(1996), which has the same characteristics of simplicity and speed as the other
merging tree algorithms, but also presents some major advantages.
Since it is based on an actual realisation of the initial density
field, it is much closer to the spirit of N--body simulations, thus allowing 
direct comparisons with numerical results, and it also seems to take into
account the spatial correlations of density fluctuations (Nagashima \&
Gouda 1997). 
Moreover, no specific collapse times are imposed {\it a priori}, and halos
form with a continuous spectrum of masses, and a variety of (Lagrangian)
shapes.  
Also the spatial information about the relative location of halos is retained
by construction, thus allowing to study their clustering properties. 
While in the original paper, the authors only discuss the halo mass
function in the case of 
scale-free power spectrum, here we consider the more realistic standard CDM 
(SCDM) cosmology (see Sec. 3).  
Moreover, we test the model reliability also in terms 
%of the halo mass function, but also by verifying the reliability 
of the mass distribution of progenitor and child halos, 
%its parent halo distributions, 
the behaviour of the largest progenitor mass as a function of redshift, 
and the probability distribution of formation times.
%redshift distribution of largest progenitors, 
%and the halo two--point correlation function. 
For the first time for a semi--analytic merging tree, 
also the halo two--point correlation function is computed, and we test it
against theoretical predictions. 
We outline the method in Sec. 2, define these quantities and compare them to
the analytic predictions in Sec. 3.
Discussion and conclusions are presented in Secs. 4 and 5, respectively.

\section{The Algorithm}

\subsection{Basic Principles}
At an `initial time' $t_{\rm i}$, consider the density field $\rho({\bf x},
t_{\rm i})$ of the Universe characterised by a mean value
$\rhoav(t_{\rm i})$, and 
small perturbations $\delta({\bf x}, t_{\rm i})= \rho({\bf x},
t_{\rm i})/\rhoav(t_{\rm i})-1$.  Through gravitational instability,
the amplitude of density fluctuations start growing proportionally
to a linear growth factor $D(t)$, i.e., $\delta({\bf x},t)=\delta({\bf
x},t_{\rm i})\!\times\!D(t)/D(t_{\rm i})$.
% Given an initial density field $\rho({\bf x}, t_{\rm i})$, its small
% perturbations 
% $\delta({\bf x}, t_{\rm i})= \rho({\bf x}, t_{\rm i})/\rhoav(t_{\rm
% i})-1$ (where $\rhoav$ is 
% the mean density of the Universe at the ``initial'' time $t_{\rm
% i}$) grow in 
% amplitude proportionally 
% to a linear growth factor: $\delta({\bf x},t)=\delta({\bf x},t_{\rm i})\,
% D(t)/D(t_{\rm i})$, where $D(t)\propto a(t)$, and $a(t)$ is the
% expansion factor 
% of the Universe.
Such a linear growth law strictly holds only when perturbations are much
smaller than unity, but it is useful to extrapolate it also into the
non-linear regime.  
In fact, the `Top-Hat' model (Gunn \& Gott 1972) shows that the
formation of 
a bound virialised object of mass M occurs at the time $\tform$ when the
density contrast of a spherical region in the initial density field, smoothed
at a scale M, reaches a critical value $\deltac$. 
This in turn corresponds to a value $\deltac^{\rm lin}(\tform)$ of the density
field {\it linearly} extrapolated to that time, or to a value
$\deltac^{\rm lin}(t_0)$ if the extrapolation is carried on until the present
epoch $t_0$.
For an Einstein--de Sitter Universe, $D(t)\propto (1+z)^{-1}$,
$\deltac\simeq 178$, $\deltac^{\rm lin}(\zform)\simeq 1.686$, and
$\deltac^{\rm lin}(z=0)\simeq 1.686\,(1+\zform)$. 
It is therefore sufficient to know the values of the density field linearly
extrapolated to $z=0$, for determining the formation epochs of DM halos. 
% collapse of a spherical 
% region in the initial density field occurs at the time $\tform$ when
% its density contrast reaches a certain critical value $\deltac(\tform)$,
% which corresponds to a value $\deltac^{\rm lin}(\tform)$ of the density field
% linearly extrapolated to that time. 
% In particular, for an Einstein--de Sitter (EdS) Universe, it results
% $\deltac(\tform)\simeq 180$, and $\deltac^{lin}(\tform)\simeq 1.686$.
% If the initial density field is linearly extrapolated until present time, 
% this corresponds to $\delta^{lin}_0=1.686\,(1+\zform)$ [in fact, $a(t)\propto
% (1+z)^{-1}$ in an EdS cosmology]. 
% It is therefore evident that knowing the value of the density contrast of a
% given region linearly extrapolated to $z=0$, immediately permits to
% obtain the collapse redshift of that region simply as: $1+\zform =
% \delta^{lin}_0/1.686$. 

Because this is the approach we adopt in the paper, we choose henceforth to
change the notation and we denote
the density field linearly extrapolated to $z=0$ as $\delta$.
Therefore, $\deltac=1.686$, and the collapse redshift of halos is 
$\zform = \delta/\deltac -1$.

\subsection{The Method}
The MCM is based on an actual realisation of the density field, 
%$\delta$. This is 
obtained with a standard initial
condition generator, by Fourier transforming waves of random phase and
amplitude drawn from a Gaussian distribution of zero mean, and variance given
by the chosen power spectrum.

A value of the density contrast $\delta$ is assigned to each of the L$^3$
{\it base cells} ($bcs$) composing a periodic cubic box of side L (for
simplicity, L=2$^l$, where $l$ is a positive integer). 
Density fluctuations are then averaged within cubic {\it blocks} of side 2,
4, 8, ..., L. At each of these smoothing levels, a set of 8 overlapping grids
displaced one relative to another by half a block length in each 
coordinate direction is used. This ensures that the density peaks are always 
approximately centred within one of the blocks in each smoothing hierarchy. 
At this point, one has a total of (15\,L$^3-8)/7$ base cells and cubic 
blocks, with side ranging from 1 to L, mutually overlapping in the
volume of the box. Each of them is characterised by a value of the density
contrast $\delta$.

All base cells and blocks are then ordered in a single list in terms of
decreasing $\delta$ (or, correspondingly, decreasing collapse redshift). 
The largest value of $\delta$ in the list fixes the earliest $\zform$ of
the realisation. 
It usually corresponds to a base cell, which thus becomes the first collapsed
object (i.e., the first {\it halo}). 
All the elements of the list are then analysed one after the other from early
times to the present, and the specific base cell or cubic block under
investigation is called {\it investigating region}. 
Whether the investigating region can collapse and
give rise to a new halo or not is decided by the following rules: 
\begin{enumerate}
\item an investigating region that does not overlap with any other 
pre-existing halo collapses and forms a new halo; 

\item if there exist two halos, each of them containing half of the
investigating  
region, the latter cannot collapse. This is to avoid the formation of very
elongated structures in linking together adjacent halos without the collapse
of any new matter. If instead there exists only one halo containing half (or
more) of the investigating region, the latter collapses and merges with it,
thus forming a new halo; 

\item after taking into account condition (ii), if the investigating region
overlaps with at least half of one (or more) pre-existing halo(s), it
collapses and merges with it (them), thus forming a new halo.
\end{enumerate}

Note that in the Lagrangian space of initial conditions, mass and volume are
equivalent quantities (M=$\rhoav$ V), and `merging' together the
investigating region with 
one or more pre--existing halos does not mean summing up their masses.
Instead, the mass (volume) of the new resulting halo is that of the old ones
plus the fraction of the investigating region that does not overlap with any
already pre--existing object.
%(nor those with which the investigating region
%merges, nor those not satisfying the merging criteria). 

Thanks to the use of overlapping grids and merging criteria, halos of a large
variety of shapes and masses are obtained. 
% spectrum of their values is also close to continuous. 
The model also contains information on the relative locations of halos, since
their positions within the box are known, and the effects of discretisation
are expected to be smaller than in the block model.
Moreover, because halo formation times are given by the density
contrasts in the list, they span a continuous range of values. 
As a drawback, the overdensity of the investigating region (which fixes the
collapse redshift of the new forming object) is not necessarily equal to the
mean overdensity ($\overline{\delta}_h$) of the resulting halo. 
Therefore, the position of an object in the merging tree occasionally
differs from that predicted by the linear theory, i.e., the assigned $\zform$
is not exactly equal to $\overline{\delta}_h/\deltac\,-1$, as it should (see
Sec. 3.2 in RT96). 
Finally, the `linking' and the `overlapping' conditions [criteria $(ii)$ and
$(iii)$, respectively] are reasonable but arbitrary, and different choices
would result in different mass functions, as discussed by Nagashima \&
Gouda (1997). 
Yet, it is not clear which are the `best' criteria. Thus, we will adopt the
original conditions $(ii)$ and $(iii)$ throughout the paper.

\section{Tests of the algorithm}
The MCM is based on the linear theory of growth of density
fluctuations, and it uses simplified criteria to describe the formation and
merging history of DM halos.
It is therefore necessary to test its reliability by comparing its results
against those of N-body simulations that directly take into account the
gravitational interactions between DM particles, and are much more realistic
in following the dynamics in the non--linear regime.

The model is required to correctly describe not only the population of halos
at a given redshift, but also how this population evolves with time.
%We have therefore computed the mass distribution of halos at given redshifts,
%their conditional mass functions (that describe the population of
%progenitors at any $z$ given that of their child halos at a later time,
%and viceversa), the behaviour of the largest progenitor mass as a function of
%redshift, and the typical formation epoch for objects with a given mass.
%For almost all these quantities, analytic predictions from the PS theory and
%its extensions exist, which are in good agreement with the results of 
%N-body simulations (e.g, Tormen 1998, and references therein).
%These formulae are therefore used as a comparison to test the MCM, and
%results are presented in next subsections.  
%Finally, we also have verified the reliability of the information about the
%relative location of halos in the MCM, by computing the halo
%auto--correlation function and comparing it to the corresponding analytic
%predictions, which in turn are also in good agreement with results of N--body
%simulations (e.g., MW96; CLMP; CPL99; Jing 1999).

As a first test, the cumulative and the differential mass functions in the
case of a scale--free power spectrum  with spectral index $n=0,$ and $-2$ have
been computed and compared to those in the original paper (Figs. 4 and 5 in
RT96). A remarkable agreement has been found. 

Here we consider the SCDM cosmology, and perform several tests against
the available analytic formulae, to verify the reliability of the model
results.  
We set the Hubble constant to H$_0 = 100\,h\,{\rm km\,\, s}^{-1} {\rm
Mpc}^{-1}$, $h=0.5$. The total and baryonic density parameters are 
$\Omega_0=1$ and $\Omega_{\rm b}=0.05$ respectively, while that corresponding
to the cosmological constant is $\Omega_{\rm\Lambda}=0$. 
The transfer function of Bardeen at al. (1986) is
adopted, and the power spectrum is normalised so that the mass variance on
scale $8\,h^{-1}$ Mpc is equal to $\sigma_8=0.67$. 

For a $256^3$ base cell realisation in a cubic box of L=100 Mpc
side (i.e., $50\,h^{-1}$ Mpc, and a total mass of about
$3.5\!\times\!10^{16}\,h^{-1}\msol$), the base cell mass is of about
$2\!\times\!10^{9}\,h^{-1}\msol$, 
and the resulting most massive halo typically has a 
mass of about $5\!\times\!10^{14}\,h^{-1}\msol$ (but see Sec. 3.1). 
The CPU time on a 500 MHz DEC Alpha workstation is only roughly 25
min, and typically 700 MB of RAM memory are required.

In the following sections, results averaged over 10 different realisations
are presented, and error bars correspond to the standard deviations of the 10
run sample. 

% We obtain a total of more than 6 millions halos, in the mass
% range from $4\,10^{9}\msol$ to $10^{15}\msol$, and the formation redshifts
% range between $z=22$ and $z=0$. 

%Anyway, only objects with masses ranging from about $5\,10^{10}\msol$ to
%about $5\,10^{13}\msol$ are reliable, because of the reasons discussed in
%next Sections.

\subsection{Mass Function}
The differential mass function of halos, is defined as the comoving number
density of halos with mass in the range [M, M+dM] at redshift $z$.
This is shown for logarithmic mass interval in the histrograms 
of Fig.~\ref{dndM} (left--hand panels), for $z=0$ and $z=3$. 
%has been computed.
% This is shown per logarithmic mass interval and for two redshifts ($z=0,3$)
% in the histograms of Fig.~\ref{dndM} (left panels), as a function of M in
% solar masses. On the top of each panel, masses in units of number of
% composing base cells are also given for reference. 

Also shown for comparison as dotted lines are the corresponding predictions
of the PS theory (e.g., LC94): 
\begin{equation}
{\der n\over{\der\,\ln\,\mass}}(\mass, z)= \sqrt{2\over\pi} \,\, \rhoav\,\,
{\deltac(z)\over{\sig2}} {\left | 
{\der\,\sigma}\over{\der\,\mass} \right |}\,\espon
{\left[-{{\deltac^2(z)}\over{2\,\sig2}} \right]},
\end{equation}
where $\deltac(z)=\deltac\!\times\!(1+z)$, and $\sigma(\mass)$ is the mass
variance of the linearly extrapolated (to $z=0$) density field smoothed on
scale M. 
In this paper, $\sigma(\mass)$ is always computed by a fitting formula
analogous to that proposed by White \& Frenk (1991), with
errors smaller than $8\%$ on mass scales ranging from
$10^{9}\msol$ to $10^{15}\msol$.

Right panels show the cumulative mass fraction for the same redshifts,
i.e. the fraction of the total mass which is in halos of mass  
above M, at redshift $z$.
% (thick solid lines correspond to the model results,
%and dotted curves to the PS predictions).  

At $z=0$ the overall agreement is good over a large range of masses, but a lack
of objects at the two ends of the mass function is evident.
At small masses, the problem seems to be inherent to the method, since 
this is also the case for scale--free power spectra (RT96; Nagashima \& Gouda
1997). 
This is a drawback of the model, which limits the reliable dynamical
range, and it is probably due to the adopted criteria for the 
formation and merging of halos. 
In fact, as detailed in Sec. 2, the ``bricks'' for the 
construction of halos are base cells and blocks composed by $8^i$ $bcs$ 
($i=1,2,..$). Thus, an object of less than 8 $bcs$ can only result if an {\it
investigating region} partly overlaps with a pre--existing halo, but does
not merge with it, and the fraction of its non--overlapping volume (which
gives rise to the new halo) is less than 8 $bcs$. In practice, the model
requires that some ``particular'' 
conditions happen in order to form halos of mass between 2 and 7 $bcs$, thus
explaining the underproduction of this kind of objects in the resulting mass
function.  
The lack of high--mass structures instead, is partly inherent to the method,
partly due to a statistical fluctuation (in different realisations in fact
the problem is more or less severe). 
At high redshift the model always tends to produce a larger number of
intermediate mass halos, and less massive objects than predicted by the PS
theory. 

These discrepancies appear to be even more severe if compared to results of
N--body simulations. It has been recently shown, that the PS mass
function already tends to predict fewer high--mass halos, and more low mass 
objects than those found in the simulations (e.g., Jain \& Bertschinger
1994; Gross et al. 1998.; Somerville at al. 1998; Tormen 1998; LS99; Sheth
\& Tormen 1999). 
An analytic formula which better agrees with numerical results has been
obtained by LS98, based on a nonspherical model for the
collapse of a perturbation, in the frame of the Zel'dovich approximation 
(but see also Monaco 1995, 1997a,b; Audit, Teyssier \& Alimi 1997; Bond \&
Myers 1996; Sheth \& Tormen 1999; Sheth, Mo, \& Tormen 1999).  

In this formalism, the displacement of a particle due to the surrounding
density field, is simply computed from the perturbation potential $\Psi$
generated by the distribution of particles in the {\it initial} conditions. 
The mass density can therefore be expressed as a function of the three 
eigenvalues of the deformation tensor (defined as the second derivative of
$\Psi$), and a virialised bound object forms when the
smaller one ($\lambda_3$) is positive. 
The idea is therefore to substitute the collapse condition of the
spherical Top--Hat model ($\delta=\deltac$), with an analogous one for
$\lambda_3$: a DM halo of mass M forms when this eigenvalue reaches a
critical value $\lambda_{3{\rm c}}$ in a region of the linearly 
extrapolated density field, smoothed on a scale M.
The resulting mass function is:
\begin{equation}
{\der\,n_{LS}\over{\der\,\ln\,\mass}}(\mass,z)={
 {{25\sqrt{10}}\over{2\sqrt\pi}} {\rhoav\over\mass}
 {\left|{{\der\,\ln\,\sigma}\over{\der\ln\mass}}\right|}
 {\lamdac\over\sigma} \times f[x],}
%f\left [\lamdac\over\sigma\right],}
\end{equation}
with $x=\lamdac/\sigma$, and:
%$f[x]$ given by:
\[
f[x] = \left[{{5\over 3}x^2} -{1\over{12}}\right]
\exp\left[-{5\over 2}\,x^2 \right]
\erfc\left[\sqrt{2}\,x \right] + 
\]
\[
\hskip 1.6truecm +\,\, {\sqrt{6}\over 8} 
\exp\left[ -{15\over 4}\,x^2 \right ] 
\erfc\left[{\sqrt{3}\over 2}\,x \right] +
\]
\[
\hskip 1.6truecm -\,\, {5\sqrt{2\pi}\over{6\pi}}\,x\,\, 
\exp\left[ -{9\over 2}\,x^2 \right ], 
\]
where $\sigma=\sigma (\mass)$, erfc($x$) is the complementary error
function, and the critical value for $\lambda_3$ has been
empirically chosen to be $\lamdac=0.37\,(1+z)$.  

Figure~\ref{dndM} shows that in comparison to the LS98 mass function (solid
curves), the MCM presents 
%In order to better quantify the extent of deviations between the MCM and
%N--body results, the LS98 mass function is shown in Fig.~\ref{dndM} as solid
%lines. 
an excess of small objects, and a significant underproduction of
high--mass halos, especially at high redshift. 

\begin{figure}
\vspace{22 pc}
\caption{{\it Left panels}: comoving number density in units of Mpc$^{-3}$, per
logarithmic mass interval, of halos of mass M at redshifts $z=0$ ({\it
upper panel}) and $z=3$ 
({\it lower panel}), as a function of M/$\msol$.
Masses are also shown in units of base cells on the top of the figure. 
Results averaged over 10 realisations of the MCM are displayed in the
histograms. Error bars show the standard deviations of the 10 run sample. 
The Press \& Schechter (1974) predictions [eq.(1)] and the Lee \& Shandarin
(1998) mass function [eq.(2)] are plotted as 
{\it dotted} and {\it solid} curves, respectively. 
{\it Right panels}: cumulative mass fraction of halos with mass larger than M
at redshifts $z=0$ and $z=3$, as a function of M. 
Results from the MCM ({\it thick solid} line) are compared to the PS and the
LS98 predictions ({\it dotted} and {\it solid} curves, respectively).
}
\label{dndM}
\end{figure}

\subsection{Conditional Mass Functions}
In this section we analyse how the population of halos identified at a given 
time has changed with respect to a different epoch.

Figure ~\ref{dfdmp} shows the mass fraction of halos of mass $\mson$ at
redshift 
$\zson$, that has already settled at redshift $\zprog$ in progenitors with
masses between $\mprog$ and $\mprog+\der\mprog$. 
Child halos have been selected at $\zson=0$ and have masses $\mson$ in the
range $\mson/\msol= \rangenum$, where, from the top to the
bottom panel, $i=11, 12, 13, 14$. The mass distribution of their progenitors
at $\zprog=1$ is shown in the left panels, that for $\zprog=3$
is plotted on the right--hand panels.
%In each panel, $\nson$ is the number of halos found at $z=0$, in the
%corresponding mass range. 

The analytic prediction for the distribution of progenitor masses is given by
(e.g., Bower 1991, Bond et al. 1991):
\[
{\der f\over{\der\,\ln\,\mprog}}(\mprog, \zprog | \mson, \zson )= 
{\mprog\over{\sqrt{2\,\pi}}} {{\deltacp-\deltacs}\over{(\sigp-\sigs)^{3/2}}} 
{\left |{\der\,\sigp}\over{\der\,\mprog} \right |}
\]
\begin{equation}
\hskip 4truecm \times\,\, \espon{\left[-{{(\deltacp-\deltacs)^2}\over{2\,(\sigp-\sigs)}} \right],}
\end{equation}
where $\delta_{{\rm c}j}=\deltac(z_j)$, $\sigma_j=\sigma(\mass_j)$.

Because numerical results shown in Fig.~\ref{dfdmp} have been obtained for
ranges of $\mson$, the analytic formula has been computed in two different
ways for an accurate comparison. 
In the MCM, the mass of a halo corresponds to the number of base cells
that compose it. Thus, when expressed in these units, the mass $\mson$
can only assume $\nvalm$ integer values in the 
range $\crange$, with $\nvalm=(\msup-\minf+1)$: i.e.,
$\mson=\{\msonk=\minf+{\it k}-1,\,\,{\it k}=1,\,\nvalm  \}$.   
Given the number $N(\msonk)$ of child halos with mass equal to $\msonk$ for
each of its possible $\nvalm$ values, the mean weighted mass in the range is:
\begin{equation}
%\avmson ={ {\sum_{\minf}^{\msup} N(\mson)\,\,\mson} \over{\rm{N_0} } }, 
\avmson ={ {\sum\limits_{k=1}^{\nvalm} N(\msonk)\,\,\msonk}\over{\nson} },
\end{equation}
where $\nson$ is the total number of halos in the chosen range of $\mson$. 
Dotted curves in Fig.~\ref{dfdmp} show results from eq.(3) computed for
$\mson=\avmson$ (whose numerical values are listed in the figure caption).
%(from top to bottom: $\avmson\simeq 2.7\!\times\!10^{11}$,
%$2.5\!\times\!10^{12}$, $2.3\!\times\!10^{13}$, $1.5\!\times\!10^{14}
%\msol$).
The abrupt fall down of dotted curves occurs at values of 
$\mprog$ near $\avmson$, because the progenitor mass obviously
cannot be larger than that of its child halo.
Since $\avmson$ is lower than $\msup$ in each panel, this
explains why dotted curves are not as extended in $\mprog$ as the
histograms are.

For a better comparison between numerical and analytic results, we
have also computed the 
%eq.(2) $\nvalm$ times (for $\mson$ equal to each
%of its possible $\nvalm$ values in the range), and we have obtained an 
{\it average} progenitor mass distribution by summing up the
$\nvalm$ single--mass distributions, each weighted with the fraction of mass
in halos of mass $\msonk$ [$N(\msonk)\times\msonk$]: 

\begin{equation}
% { \langle{\der f\over{\der\,\ln\,\mprog}}\rangle ={ { 
% \sum_{\minf}^{\msup}
% { {\der f(\mson)} \over{\der\,\ln\,\mprog} }\,N(\mson)\,\mson }
% \over{ \sum_{\minf}^{\msup} N(\mson)\,\,\mson } } 
% ,}
%\langle{\der f\over{\der\,\ln\,\mprog}}\rangle ={
\left <{\der f\over{\der\,\ln\,\mprog}}\right > ={
 {\sum\limits_{k=1}^{\nvalm}\,f(\msonk)\,N(\msonk)\,\msonk } \over{
 \sum\limits_{k=1}^{\nvalm} N(\msonk)\,\,\msonk } },
\end{equation}
where $f(\msonk)=(\der f/\der\,\ln\,\mprog)$ is the single--mass progenitor
distribution for child halos of mass $\msonk$ as given in eq.(3). 
$\mprog$ is obviously also a function of 
$\mson$, and it is required that $\mprog\le\mson$. 
The average progenitor mass distribution is plotted in Fig.~\ref{dfdmp} as
solid curves. 
By construction, $\langle{\der f/{\der\,\ln\,\mprog}}\rangle$ is a sum of
curves of the same kind of the dotted lines, with the sharp cut--off at 
progenitor masses very similar or equal to that of their child 
halos. This is the reason for the oscillations in the solid curves for
values of $\mprog$ between $\minf$ and $\msup$. 

A lack of objects with masses between 2 and about 7
$bcs$ is apparent, just as was the case for the mass function. 
The MCM also appears to systematically underproduce progenitors with
mass similar to that of their child halos. 
%This difference is larger when large values of $\mson$ are
%considered (bottom panels), because of the inherent lack of high mass objects
%in the MCM discussed in the previous section. 
At intermediate masses, an overall good agreement between the MCM results
and the analytic predictions is found, with a possible slight overproduction
of halos in the model. 
When compared to N-body simulations, these discrepancies may become 
more severe, since simulations appear to have fewer/more halos than predicted
by EPS theory in the intermediate/high mass range (Somerville et al. 1998;
Tormen 1998). 
For the less massive child halos (top panels), numerical and analytic
results only agree over a small range of $\mprog$.
This is due to the lack of low mass objects, and to the
fact that the minimum mass in the model is limited to 1 base cell, thus 
not allowing to accurately follow back in time the past history of small
halos. 

\begin{figure}
\vspace{11pc}
\caption{Progenitor mass distribution at redshifts $\zprog=1$ ({\it left
panels}), and $\zprog=3$ ({\it right panels}), for child halos in four
different mass ranges $\mson$ at $\zson=0$ ({\it from top to bottom}:
$\mson/\msol=[10^i,10^{i+1}[\, ,i=11, 12, 13, 14$). 
Average results from 10 realisations of the MCM are displayed in the
histograms, and their standard deviation is shown as error bars.
The average number of child halos found in each mass range is, from top
to bottom: $\nson=34839, 5545, 822,$ and 67.
{\it Dotted} curves correspond to the progenitor distribution computed from
eq.(3) for $\mson$ equal to the mean mass in the corresponding range
($\mson=\avmson\simeq 2.7\!\times\!10^{11}$, $2.5\!\times\!10^{12}$,
$2.3\!\times\!10^{13}$, and $1.5\!\times\!10^{14}\,\msol$). 
{\it Solid} curves are the average distribution in each mass range, computed
from eq.(5). }
\label{dfdmp}
\end{figure}

The reverse conditional probability that a halo of mass $\mprog$ at $\zprog$
is incorporated at a later time $\zson$ in a halo of mass between 
$\mson$ and $\mson+\der\mson$, 
%has also been computed from the MCM. It 
is shown per logarithmic mass interval in the histograms of
Fig.~\ref{dfdms}.
%, as a function of $\mson/\msol$. 
Progenitors of mass $\mprog/\msol=\rangenum$, with $i=11, 12, 13$, are
selected at $\zprog=1$ (left panels) and $\zprog=3$ (right panels), and the
mass distribution is computed for their child halos at redshift zero. 
No results for progenitors with masses between $10^{14}$ and $10^{15} \msol$
are shown, because too few of them have already formed at redshift 1 and 3.
%(actually, also results in the last panel are not reliable, because only 8
%progenitors in this mass range are found at $\zprog=3$). 

Given all the objects of mass $\mprog$ at $\zprog$, the analytic prediction
from the EPS theory for the mass distribution of their
child halos at $\zson$, when expressed per mass logarithmic interval, is
given by (e.g., LC94):  
\[
{\der f\over{\der\,\ln\,\mson}}(\mson, \zson | \mprog, \zprog )= 
{\mson\over{\sqrt{2\,\pi}}} {{\deltacs\,(\deltacp-\deltacs)}\over\deltacp}\,\,
{\left |{\der\,\sigs}\over{\der\,\mson} \right |}
\]
\begin{equation}
\hskip 1.2truecm \times 
{\left[{\sigp\over{\sigs(\sigp-\sigs)}}\right]^{3/2}} 
\espon{\left[-{{(\deltacs\,\sigp-\deltacp\,\sigs)^2}\over
{2\,\sigp\sigs\,(\sigp-\sigs)}} \right],}
\end{equation}
where the notation is the same as in eq.(3). 

As before, eq.(6) has been computed in two different ways, in order to get 
an accurate comparison with the numerical distributions.  
Results for $\mprog=\avmprog$, the mean weighted progenitor mass in the range
$\crange$ (analogous to $\avmson$), are plotted in Fig.~\ref{dfdms} as dotted 
lines.
A sharp cut off occurs for values of $\mson$ near to $\avmprog$,
because child halos cannot be less massive than their progenitors. 
Since $\avmprog$ is larger than $\minf$ in each panel, this
explains why dashed lines are not as extended in $\mson$ as histograms are.
A more appropriate comparison between numerical and theoretical results is
obtained if the {\it average} child mass distribution $\langle{\der
f/{\der\,\ln\,\mson}}\rangle$ is considered, instead of that relative to
progenitors with mean mass $\avmprog$: $\der f(\avmprog)/{\der\,\ln(\mson)}$. 
The computation of $\langle{\der f/{\der\,\ln\,\mson}}\rangle$ is analogous 
to that in eq.(5), and results are plotted in Fig.~\ref{dfdms} as solid lines.
%The lack of high mass halos discussed in Sec.3.1. is once more apparent, but 

An overall agreement is found between MCM results and the EPS
theory predictions, that in turn fit reasonably well N-body simulations
(LC94). 
However an oscillating behaviour of the child halos mass distribution can be
recognised in the histograms. Actually it is more evident
when a different binning is used (here results are binned on a mass grid
$\mson=2^i\,bcs,\,i=0,1,2..$), and it seems inherent to the method. Also the
halo mass function and the progenitor distribution present analogous features,
and oscillations appear to occur with peaks corresponding to the block 
masses of $8^i, i=1,2,.. \,bcs$, and with troughs in between. 
Moreover, the same trend is found in the halo mass function for the scale
free power spectrum (in particular for the spectral index $n=-2$, that fits
the CDM spectrum over a significant range of masses; see RT96). 

\begin{figure}
\vspace{11pc}
\caption{Mass distribution of child halos at redshifts $\zson=0$, given
the progenitors at $\zprog=1$ ({\it left panels}), and $\zprog=3$ ({\it right
panels}) with masses, $\mprog/\msol=[10^{i},10^{i+1}[\, , i=11, 12,
13$, from {\it top to bottom}. 
Average results from 10 realisations of the MCM are plotted as histograms,
and their standard deviation is shown as error bars.
The number of progenitors found in each mass range at $\zprog=1$ is, from
top to bottom: $\nprog=59898, 6696,$ and 431, while at $\zprog=3$,
$\nprog=52213, 1724,$ and 8.
{\it Dotted} curves refer to the children mass distribution computed from
eq.(6) for the mean progenitor mass in the corresponding range (from top to
bottom): 
$\avmprog\simeq 2.6\!\times\!10^{11}, 2.2\!\times\!10^{12},$ and
$1.7\!\times\!10^{13} \msol$ for progenitors at $\zprog=1$; $\avmprog\simeq
2.3\!\times\!10^{11}, 1.9\!\times\!10^{12},$ and  
$1.5\!\times\!10^{13} \msol$ in the right panels. 
{\it Solid} curves are the average distribution in each mass range, computed
in the same way as in eq.(5), as detailed in the text.
}
\label{dfdms}
\end{figure}

\subsection{Largest progenitor history}
By analysing the variation with redshift of the largest progenitor mass, 
information can be obtained on how halos build up in time, 
%mergers occur during halo evolution, i.e., 
whether they preferably form via a continuous and slow accretion of small
objects, or whether their mass suddenly increases because of nearly equal--mass
merging events, or by mergers of several sub--units at the same time. 
%The way halos assemble has relevant consequences on the final galaxy
%population within them, thus it is important that the MCM histories agree
%with those obtained from N--body simulations. 
A different behaviour is expected for halos of different masses, with larger
objects preferably assembling at recent epochs, and smaller halos showing a
more delayed and smooth evolution with time. This is shown for instance, in
KW93, both from their merging tree model, and from N--body
simulations (their Figs. 5 and 6, respectively).

We have looked at the past history of halos with current mass $\mson=
10^{11}, 10^{12}, 10^{13}, 10^{14} \msol$, randomly selecting 30 objects for
each value of $\mson$ to show the scatter in the merging histories.
The ratio between the mass of the largest progenitor M$_1$, and that of its
child halo $\mson$ is plotted in Fig.~\ref{largp}, as a function of $1+z$. 
%Each panel corresponds to one of the four values considered for $\mson$ (see
%labels). 
%Largest progenitors have been searched on a grid of 20 redshifts between $z=0$
%and $z=6$, with the constant step in log(1+$z$).
%The not smooth (?) trend of curves in the first panel once more shows that the
%model can not follow in an accurate way the past history of $10^{11}\msol$
%halos, because of the limit to a minimum of 1 base cell for the mass
%resolution.  
For all the masses, the expected trends are obtained, with larger
halos preferably assembling through major mergers at low $z$, and smaller
objects gradually forming in a smoother way by accreting small mass objects
over a larger interval of time. 
A qualitative good agreement of both trends and scatters is also found
between the present results and those of KW93. Moreover, halo collapse occurs
at more recent epochs here, as expected when a SCDM cosmology is considered
instead of an open model ($\Omega_0=0.2$ in KW93). 

% By looking at the time when the largest progenitor has for the first time a
% cd ..mass equal to at least $\mson/2$, we have also derived the mean and the 
% median formation redshifts of child halos. They respectively are equal to
% 1.16 and 1.1 for $\mson=10^{11}$, 1.04 and 1.1 for $\mson=10^{12}$, 0.53 and
% 0.37 for $\mson=10^{13}$, ?? and ?? for $\mson=10^{14}$, in good  
% agreement with the values quoted in the previous section.[?? spiegare perche'
% sono diversi..]

\begin{figure}
\vspace{11pc}
\caption{History of the most massive progenitor of 30 halos selected at
$\zson=0$.
The $y$--axis represents the ratio of progenitor mass $\mprog$ to final halo
mass $\mson$. 
%Redshift variation of the ratio between the mass of the largest
%progenitor M$_1$ and that of its child halo $\mson$ at $\zson=0$. 
The four panels refer to four different values of $\mson/\msol: 10^{11},
10^{12}, 10^{13}, 10^{14}$.
%In order to show the scatter between different merging history, 30 child
%halos have been randomly selected for each mass $\mson$. 
}
\label{largp}
\end{figure}

\subsection{Formation redshift}
In the hierarchical clustering scenario, massive halos form by
accretion of lower mass structures. Therefore, their formation redshift is
expected on average to be lower than that of small objects.
Actually, because of the continuous evolution in mass due to the hierarchical
nature of the process, the definition of `halo formation time' is
not straightforward.  
In this paper, we adopt the definition of LC93, as the time when half the
mass of the halo is assembled, i.e., when a progenitor with mass
equal to half or more that of its child halo appears for the first time. 

% Following this definition, we have computed the formation redshifts $\zform$
% for halos having mass $\mson=[10^{i},\,5\!\times\!10^{i}]\,\msol$,
% $i=1,2,3,4$ at $\zson=0$. 
% Their differential probability distribution is plotted in the 
% histograms of Fig.~\ref{zform} as a function of $(1+\zform)$. Each panel
% refers to one of the four mass ranges, and the mean number (averaged 
% over the 10 realisations) of halos found with that mass at $\zson=0$ is also
% labeled as $\nson$. 
%The redshift grid used for this computation has constant step in log(1+$z$).
Figure ~\ref{zform} shows the distribution of formation redshift for halos with
mass $\mson=[10^{i},\,5\!\times\!10^{i}]\,\msol$, $i=11,12,13,14$ at $z=0$.
In agreement with results of the previous section, high mass objects
tend to form at more recent epochs, while lower 
mass halos typically collapse earlier and over a larger interval of time.  
The mean formation redshifts for halos in the four mass ranges, from lower to
higher $\mson$, are: $\avzform=1.55, 1.03, 0.66, 0.46$. 

As discussed in LC93 and LC94, the probability that a halo of mass $\mson$ at
redshift $\zson$ has a progenitor with mass between $\mson/2$ and $\mson$ at
$\zprog$, gives the probability that its formation epoch was earlier than
$\zprog$. In differential form, the probability distribution of formation
redshifts is therefore given by: 

\begin{equation}
{\der p\over{\der\zform}}(\zform | \mson,
\zson)={ \int_{\mson/2}^{\mson}{ {\mson\over\mprog}\, {\left[
{\partial\over{\partial\zform}} {\left({\der {\it f}\over{\der\mprog}}
\right)}\right]}\der\mprog }},
\end{equation}
where ${\der f/\der\mprog}={\der f(\mprog,\zform | \mson,\zson)/\der\mprog}$
is the progenitor mass distribution (see Sec. 3.2).
Formation times computed by means of the previous formula are found to be in
good agreement with N-body simulation results, except for halos on cluster
scales that form earlier than predicted by the EPS theory (LC94; Tormen 1998;
note however that these conclusions are drawn for scale--free power spectra 
only).

Once again, numerical results are derived for ranges of masses $\mson$,
thus the analytic prediction for $\zform$ has been computed in the same 
way as discussed in Sec. 3.2. 
In this case however, the probability distribution for the average mass
$\der p(\zform | \avmson\,\zson)/\der\zform$, and the {\it average} probability
distribution $\langle{\der p/{\der\,\zform}}\rangle$ are almost
indistinguishable in the chosen range of $\mson$. 
Only the former is therefore shown in Fig.~\ref{zform}, where solid
curves 
correspond to eq.(7) solved for $\mson=\avmson\simeq
2.2\!\times\!10^{11},\,2\!\times\!10^{12},\,1.7\!\times\!10^{13},\,1.4\!\times\!10^{14}\msol$.
Within the error bars, a very good agreement is found for the most
massive objects.
%, but a systematic shift of the peak of their formation times
%towards more recent epochs is apparent ($z\simeq 0.24$ in the
%MCM, $z\simeq 0.33$ in the analytic prediction).
%They show an opposite trend with respect to the EPS prediction, compared to
%more massive halos, 
For intermediate mass halos, the epoch when they first appear, as well
as the rising of the probability distribution with decreasing $z$ is
well reproduced by the MCM.
However, they do not present the expected peak of formation epoch,
but instead still form at 
%tend to collapse later than predicted, and are still forming also at 
very recent times, in contradiction with the expectations of the EPS theory.
A severe disagreement is found for small objects, with the
departure of MCM relative to the EPS theory going in the opposite sense. 
Low mass halos in fact preferentially collapse and stop forming at earlier
epochs than predicted, with a peak of formation at about
$\zform=1.6$, instead of $\zform=0.85$.  
No significant improvements are obtained if different values for the collapse
threshold $\deltac$ are adopted. 
This confirms once more that the history of low mass objects is not well
followed in the model. 

\begin{figure}
\vspace{11pc}
\caption{Differential probability distribution of formation redshifts
$\zform$ for halos at $\zson=0$ with masses $\mson/\msol=
[10^{i},\,5\!\times\!10^{i}]$, $i=11, 12, 13, 14$ (see labels), as a
function of $(1+\zform$).  
Histograms and error bars result from the average over 10 realisations of the
MCM. 
$\nson$ labels the  number of halos found at $\zson=0$ in the corresponding
mass range. 
{\it Solid curves} refer to the analytic prediction of the EPS
theory computed 
from eq.(7), for $\mson=\avmson\simeq
2\!\times\!10^{11},\,2\!\times\!10^{12},\,1.7\!\times\!10^{13},\,1.4\!\times\!
10^{14}\,\msol$. 
}
\label{zform}
\end{figure}

\subsection{Two--point Correlation Function}
% Galaxies are believed to be biased tracers of the underlying mass, in a way
% that also depends on the assumed cosmology. 
% Since they form within the potential well of DM halos, studying the
% clustering properties of these latter objects can provide clues about the
% formation and the evolution of galaxies, and help discriminate between
% different cosmological models. 
Since the relative positions of halos within the box are known by 
construction, the MCM also contains information about their spatial
distribution. 
%To test its reliability, 
We have computed the two--point
autocorrelation function of DM halos, by counting the number of objects
separated by a distance $r$, and comparing it to the value expected for a
Poissonian distribution: 

% Halos centers correspond to the geometric centers of collapsed regions,
% and distances are computed in a grid of values ranging from $r=0.4\,{\rm
% Mpc}$ (the base cell size) to $r=100\,{\rm Mpc}$, with constant step
% $\Delta{\rm log}(r/{\rm L})=0.1$. 
\begin{equation}
\xi(r) = {N_{DD}(r)\over{ N_{RR}(r)}}-1,
\end{equation}
where $N_{DD}(r)$ is the number of pairs whose geometric centres are
separated by a distance between $r$ and $r+dr$, and $N_{RR}(r)$ is the same
quantity if 
halos were randomly distributed in the same volume:
$N_{RR}(r)=(1/2)$N$^2_0(dV/V)$, where $\nson$ is the total number of halos, 
$dV$ is the volume of the shell at $r$ with thickness $dr$, and $V$ is the
total volume of the box.  

Results for halos selected in four mass ranges at $z=0,\, 1$ and 3 are shown
as circles in Figs.~\ref{zf0}, Fig.~\ref{zf1} , and Fig.~\ref{zf3},
respectively. 
% Each panel corresponds to a different mass range (see labels), 
% and the corresponding mean number of halos (averaged over 10
% realizations) is also labeled as N. 
% The three vertical thick lines mark the typical Lagrangian radius
% [R$=(3\mass/4\,\pi)^{1/3}$] of halos in the given mass range: the two shorter
% ones correspond to the minimum and the maximum mass in the interval, while
% the longer one to the mean mass weighted by the mass function.
Note that points are found at separations smaller than the typical halo sizes
(marked by the vertical thick lines in plots). 
This is a consequence of the nonspherical shape of halos in the MCM, allowing
the distance between two centres to be smaller than the spherical radius R
artificially attributed to each object in this computation.  

Using an approach based on the EPS theory, CLMP give an analytic formula for
the halo two--point correlation function, which is valid 
% In their derivation they do not introduce any background scale
% R$_0$, thus the only condition for their expression to be valid is imposed
% by the spatial extension of dark matter halos: halo clustering is
% reliably described by their formula only
for separations $r$ larger than R.
% the typical Lagrangian radius R of halos.
In particular, when $r\gg$R, the correlation function of objects of mass M
identified at redshift $z$, can be expressed as: 
\begin{equation}
\xih(r,\mass,z) = {b_1^2(\mass,z)\,\xim(r,z) + {1\over 2}
b_2^2(\mass,z)\,\xim^2(r,z) + ...,}  
\end{equation}
where $\xim(r,z)$ is the matter correlation function (the Fourier transform
of the power spectrum) linearly extrapolated to redshift $z$. 
The linear bias function $b_1(\mass,z)$ was already obtained by MW96
using a different approach still based on the EPS theory. It is given by:
\begin{equation}
b_1(\mass,z) = {\deltac\over{\sigma^2(\mass,z)}} -
{1\over{\deltac}}, 
\end{equation}
where $\sigma(\mass,z)$ is the mass variance linearly extrapolated to
redshift $z$: $\sigma(\mass,z)=\sigma(\mass)\,(1+z)^{-1}$. 
CLMP show that the second order bias factor is:
\begin{equation}
b_2(\mass,z) = { {1\over{\sigma^2(\mass,z)}} {\left[
{{\deltac^2}\over{\sigma^2(\mass,z)}} -3 \right ]}. }
%b_2(\mass,z) = { {1\over{\sigma^2(\mass)}} {\left[
%{{\deltac^2\,(1+z)^2}\over{\sigma^2(\mass)}} -3 \right ]}. }
\end{equation}
If the typical nonlinear mass $\mstar(z)$ for dark matter halos is
defined as $\sigma[\mstar(z),z]=\deltac$, it results from 
eq.(10) that the first order bias vanishes for M=$\mstar$, and $\xih$ is
then determined by the second order term only. 
For redshifts $z=0,\,1,\,3$, the values of $\mstar$ are $3.4\!\times\!10^{13},
1.2\!\times\!10^{12}$, and $6.2\!\times\!10^{9} \msol$, respectively.

When a finite range of halo masses is considered instead of a single value of
M, the theoretical halo correlation functions can still be estimated by
eq.(9), with the 
two bias factors replaced by their mean values in the mass interval, weighted
by the mass function $n(\mass,z)=\der n/\der\mass$:
\begin{equation}
% \overline{b}_i = { {\int_{\minf}^{\msup}b_i(\mass,z)\, n(\mass,z)\,d\mass}
% \over{ \int_{\minf}^{\msup}\,{n(\mass,z)\,d\mass} }}  \,\,\,\,i=1,2. 
 \overline{b}_i = { {\int\limits_{\minf}^{\msup}b_i(\mass,z)\,
 n(\mass,z)\,d\mass} \over{ \int\limits_{\minf}^{\msup}\,{n(\mass,z)\,d\mass}
 }}  \,\,\,\,\,\,\,\,i=1,2.  
\end{equation}

Long dashed curves in Figs.~\ref{zf0}, Fig.~\ref{zf1}, and Fig.~\ref{zf3}
have been computed by means of eq.(9)--(12) for the corresponding mass ranges
and redshifts. 
%The values of the two mean bias factors are labeled in the figures. 
Also shown for comparison are the linear mass correlation function at each
redshift (dotted lines), and $\xih$ computed with the linear term in eq.(9)
only (dashed--dotted lines).
For all redshifts and halo masses, the autocorrelation function derived from
the MCM is in a remarkable good agreement with the
predictions of the EPS theory.
Even if the analytic formula for $\xih$ has been obtained in the limit of
separations much larger than R, a reasonable agreement is also found when
this condition is not exactly satisfied. 
Moreover, results of the MCM are well described by the linear bias
relation also in the (slightly) non--linear clustering regime (i.e., for
separations where $\xim(r)$ is slightly larger than unity), and thanks to the
second order term, eq.(9) still provides a very good description of the model
correlation function, even for masses near $\mstar$, where $b_1$ vanishes.

Such an agreement between the MCM results and analytic predictions derived from
the EPS theory only ensures the reliability of the model in correctly taking
into account the clustering of high mass ($\mass\gsim\mstar$) halos, but it
also highlights its limitations for small objects.  
Indeed, accurate comparisons with N--body simulations show that the
correlation function given by eqs.(9)--(12) correctly describes numerical
results for halos with masses larger than $\mstar$, but significantly
overestimates the clustering of small mass objects (PCL; Jing 1999; Sheth
\& Tormen 1999; Sheth, Mo, \& Tormen 1999). 
For $\mass <\mstar$, the analytic bias factor $b_1$ is significantly lower
(more negative) than that found in numerical simulations, whereas a better fit
to the N--body Lagrangian 
correlation function (with errors within the $15\%$ for a CDM cosmology) is
obtained by means of the linear term in eq.(9), with $b_1$ replaced by (Jing
1999; but see also Sheth, Mo, \& Tormen 1999):  
\begin{equation}
b_J(\mass,z) = { {\left
[{{\sigma^4(\mass,z)}\over{2\,\deltac^4}}+1\right]
%^\alpha}
^{(0.06-0.02 n)}}
\hskip -1.5truecm\times{\left [1+b_1(\mass,z)\right]} -1,}
\end{equation}
%where $\alpha=(0.06-0.02\,n)$, and 
where $n$ is the index of the power spectrum
$P(k)$, computed as: 
\begin{equation}
n = { {{d\,{\rm ln}\,P(k)}\over{d\,{\rm ln}\,k}} \vert_{k={ {2\pi}/{R}}}}.
\end{equation}

The correlation functions computed as $\xih=\bJeff^2\,\xim$ are plotted in 
Figs.~\ref{zf0}--\ref{zf3} as solid curves. 
Relative to these N--body based correlation functions, the MCM overestimates
the clustering of halos on the low mass ($\mass < \mstar$) regime.  
%In order to better quantify the extent of the deviations between MCM and
%N--body results at small halo masses, 
%is plotted in Fig.~\ref{zf0}, Fig.~\ref{zf1}, and
%Fig.~\ref{zf3} as solid curves, and the value of $\overline{b}_J$ is
%also labelled. 
%As evident, the MCM overestimate of halo clustering for small (M $<\mstar$)
%objects is significant with respect to the fitting formula. 

\begin{figure}
\vspace{11pc}
\caption{Auto--correlation function of halos with mass $\mass$ selected at
redshift $z=0$. Each panel refers to a different mass range: $\mass/\msol
=[10^{i},\,5\!\times\!10^{i}]$, with $i=11, 12, 13, 14$ (see labels). 
Average results and standard deviation from 10 MCM realisations are plotted as
{\it circles} and error bars, and the average number of halos found for each
mass range is also indicated as N.
Separations are in units of the box length (L=100 Mpc = 256 $bcs$). 
The three {\it vertical thick lines} mark the typical Lagrangian radius R of
halos in the given range of $\mass$: the two shorter
ones correspond to the minimum and the maximum mass in the range, while the
longer one refers to the mean mass in the interval, weighted by the mass
function. 
{\it Long dashed} curves have been computed by use of the linear and the second
order bias factors [Catelan et al. 1998; see eqs.(9)--(12)], and the values
of $\overline{b}_1$ and $\overline{b}_2$ are labelled in each panel.   
Also shown are the linear mass correlation function at the given redshift
({\it dotted} curves), and $\xih$ computed with the linear bias only, as first
discussed by Mo \& White (1996; {\it dashed--dotted} lines). 
{\it Solid} curves corresponds to the correlation functions computed as
$\xih=\bJeff^2\,\xim$, where the value of $\overline{b}_J$ (see label) is
derived from Jing's formula [eq.(13)], which provides a good fit to
N-body simulations. 
%The value of $\mstar$, where the 
The first order bias vanishes at $\mstar\simeq 3.4\!\times\!10^{13}\msol$.
}
\label{zf0}
\end{figure}

\begin{figure}
\vspace{11pc}
\caption{The same as in Fig.~\ref{zf0}, but for halos selected at $z=1$. 
No results for halos with mass in the range
$[10^{14},\,5\!\times\!10^{14}]\,\msol$ are shown because only 5 of them have
already formed at this epoch.
At $z=1$, $\mstar\simeq 1.2\!\times\!10^{12}\msol$.  
}
\label{zf1}
\end{figure}

\begin{figure}
\vspace{11pc}
\caption{The same as in Fig.~\ref{zf0}, but for halos selected at $z=3$. 
No results for larger masses are shown because too few, or no high-mass
halos, have already formed at this redshift.  
Here, $\mstar\simeq 6.2\!\times\!10^{9}\msol$.  
}
\label{zf3}
\end{figure}

\section{Discussion}
As far as the halo mass function, and the conditional
probability distribution of progenitor and child halos are concerned, 
a good general agreement between MCM results and PS and EPS analytic formulae
is found, but an underproduction of low-mass objects in the model is apparent. 
%Together with the less severe, but systematic underestimate of very massive
%halos, this limits the reliable dynamic range of the MCM between about 8 and
%$2.5\!\times\!10^4$ base cells ($3\!\times\!10^{10}$ and $10^{14}\,\msol$,
%for the present choice of cosmological parameter and box size). 
This limits the mass resolution of the MCM to a minimum of 8
base cells ($3\!\times\!10^{10}\,\msol$, for the present choice of
cosmological parameters and box size).  
Compared to the mass function from N-body simulations (well described when a 
nonspherical model for the collapse of density fluctuations is considered; 
see LS98; LS99; Sheth, Mo, \& Tormen 1999, and references therein), the MCM
produces a significantly lower number of high-mass halos, especially at early
times.  
Since a finite box is used for representing the Universe, the effective
amplitude of the mass variance on large scales is smaller than that expected
from the input power spectrum, used in the computation of the 
analytic formulae. Such an effect may in part be responsible for the
underproduction of high-mass halos in the MCM with respect to theoretical
predictions.  
Also changing the {\it linking} and the {\it overlapping} conditions 
(see Sec. 2.2) helps obtaining larger mass halos, especially at high $z$, but
it is not clear which are the best criteria (see also Nagashima \&
Gouda 1997). 
Moreover, oscillations in the mass functions  occurring
at block masses of $8^i$ base cells, $i=1,2, ..$ , are apparent, 
%\footnote{Similar oscillations are also present in the block model (Dos
%Santos, private comunication).}
but may possibly disappear if 
different criteria are adopted when deciding whether to merge or not
pre-existing 
halos and form a new structure, as well as if set of grids displaced in a
different way (no longer by half a block--length) are used. 
%These possibilities will be explored in a future work.

%As a general result, the past history of small halos is not well followed in
%the model, probably because of its limited mass resolution. This can be seen
%both from the progenitor mass function, and the probability
%distribution of formation redshifts.

The distribution of formation redshifts is in good agreement with
analytic predictions for high--mass halos, even if the peak of
formation is  systematically shifted towards more recent epochs. 
For intermediate mass objects, the MCM correctly reproduces the 
analytic expectations only at high redshifts, then it keeps forming halos also
at very recent times, in contradiction with the EPS theory.
Once more, a failure of the model in describing the history of
low mass halos is evident, since they systematically form at earlier epochs
than predicted. 

%If the value of the critical density is lowered in the model, 
%the history of most massive halos is better described.
%In fact, the discrepancy at the high-mass end of the mass function, is 
%less severe for $\deltac=1.5$, and the distribution of formation redshift for 
%$10^{14}\msol$ halos is in good agreement with the analytic predictions.
%A similar improvement is seen for low mass objects when
%$\deltac$ is increased, thus possibly suggesting that the threshold value for
%the collapse of density fluctuations is a decreasing function of M.
%However, the problem of a too recent formation of intermediate mass halos
%seems unavoidable.  

Finally, 
%we have also computed the DM halo correlation function and compared
%it to the available theoretical predictions. 
a remarkable agreement of the two--point correlation function is found
with respect to the predictions derived from the EPS   
theory (MW96; CLMP), for all considered masses and redshifts. 
This ensures that the model reliably retains information about the spatial
correlation of high mass halos (M$\gsim \mstar$), but it also overestimates
the clustering of small objects (as do all analytic formulae).
The amplitude of their correlation function in fact is significantly (2--3
times) higher than that predicted by the fitting formula recently proposed by
Jing (1999), that correctly describes the correlation function found in
numerical simulations. 
As discussed by Jing (1998, 1999) and PCL, this 
difference between N--body and EPS results in
Lagrangian space, suggests that the criteria adopted in
the PS theory for identifying bound virialised objects in the initial
conditions are inadequate. 
The assumption of spherical symmetry for the collapse is certainly a strong
simplification, and it also affects both the mass function and the typical
formation epoch of structures.  
Considering that halos in the MCM are produced
with a large variety of shapes, we intend to adopt a nonspherical condition 
and study its effects on the resulting mass function and formation redshift
distribution in a future paper. 

%Unless some weakness, the MCM is a useful method to 
%follow the hierarchical growth of structures, and address the issue of galaxy
%formation and evolution by means of simple prescriptions. 

The physical processes ruling gas cooling, dissipative collapse, star
formation, evolution and feedback  
(as well as interactions and merging between galaxies) are currently
implemented in merging history trees of DM halos, 
so far obtained through two main approaches. 
In semi-analytic models (KW93; Kauffmann, White \& Guiderdoni 1993;
Kauffmann, Guiderdoni \& White 1994; Cole et al. 1994;
Baugh et al. 1998; Somerville \& Primack 1999; and papers in these series),
the merging history of DM halos is built through Monte-Carlo realisations of
the block model or EPS formalism, with no or not accurate spatial information. 
More recently (Kauffmann, Nusser \& Steinmetz 1997;
Governato et al. 1998; Benson et al. 1999), DM halos have been selected from
cosmological N-body simulations, but their merging trees are still computed
with the Monte Carlo technique. 
In a ``fully'' hybrid model (Roukema et al. 1997; Kauffmann et al. 1999),
merging trees are also 
computed from the output of large N-body simulations, and as a consequence they
retain the spatial and dynamical information of the parent simulation, but
they suffer from its limited mass resolution and expensive CPU cost.

The interest of the MCM is that it represents an intermediate approach.
It is very fast, and it partly retains spatial information in the linear or
weakly non-linear regime. 
A priori, it suffers from the same resolution problem 
as merging trees built from N-body simulations. For the same choice of 
cosmological parameters and box length, the $256^3$ base cells have 
the same mass as the $256^3$ particles, and reliable halos cannot be obtained
below $\sim$ 8 base cells or 10 particles.  
However, its low cost in terms of CPU time allows to run realisations of
sub-boxes, thus improving the mass resolution.
Moreover, many choices of the cosmological parameters, shape and
normalisation of the power spectrum of linear fluctuations can be tested. 
So the MCM appears as a versatile and rapid method to test physical ideas
about galaxy, group and cluster formation in various cosmologies,
mostly when some degree of spatial information can be useful. 

In particular, the MCM can be suitable for studying galaxy clusters,
mainly at low redshifts, where a good agreement between MCM and
analytic results is found, not only in terms of mass functions,
but also in the distribution of formation redshifts, as well as in the
halo two--point correlation function.
Also the population of Lyman--break galaxies at $z=3$ can be reasonably well
studied by means of the MCM.  
In fact, these objects are often interpreted as star--forming galaxies
located at the centre of halos of about $10^{12}\,\msol$ (e.g., Steidel et
al. 1996; Giavalisco, Steidel \& Macchetto 1996; 
Steidel et al. 1998; Giavalisco et al. 1998; Baugh et al. 1998; but
see also Somerville, Primack \& Faber 1998).
For these masses and redshifts, the model provides a reasonably good
description of both the mass distribution and the formation history.
Moreover, the correlation function fairly matches the numerical results over
a large range  of halo separations, thus allowing in principle to investigate
the clustering properties of Lyman--break galaxies. 

\section{Conclusions}
The Merging Cell Model originally proposed by Rodrigues \& Thomas
(1996) for a scale--free power spectrum, has been developed in
the case of the SCDM cosmology. 
Its reliability  has been tested not only in terms of the halo
mass function, but also comparing the distributions of the progenitor
and child masses, as well as that of halo formation times, to
the analytic predictions derived by the Press \& Schechter theory and
its extensions. 

For the first time in the case of a semi--analytic merging tree model,
we have also computed the halo two--point correlation function, and
compared it to the available theoretical predictions. 

We have stressed the major successes of the model, as well as its main
weakness, and several possible solutions to improve it have been proposed.

Two main fields where the use of this
method can be of particular interest have been recognised.
It appears to be a suitable tool for studying the properties of
cluster--scale objects, mainly at low redshift, as well as the population of 
Lyman--break galaxies, and their clustering at high $z$. 

We intend to apply the method in a more realistic cosmological
scenario (as the open and the lambda CDM), and directly test it 
against N--body simulations in a forthcoming work.

\section*{Acknowledgments}
BL is very grateful to S. Colombi, S. Matarrese, L. Moscardini, C. Porciani, 
and G. Tormen for many useful discussions. 
We also thank D. Pogosyan for having kindly provided us with its code for the 
generation of initial conditions. 
BL is supported by a Marie Curie Training Grant (category 20), 
under the TMR Activity 3 of the European Community Program.

\bsp
\label{lastpage}
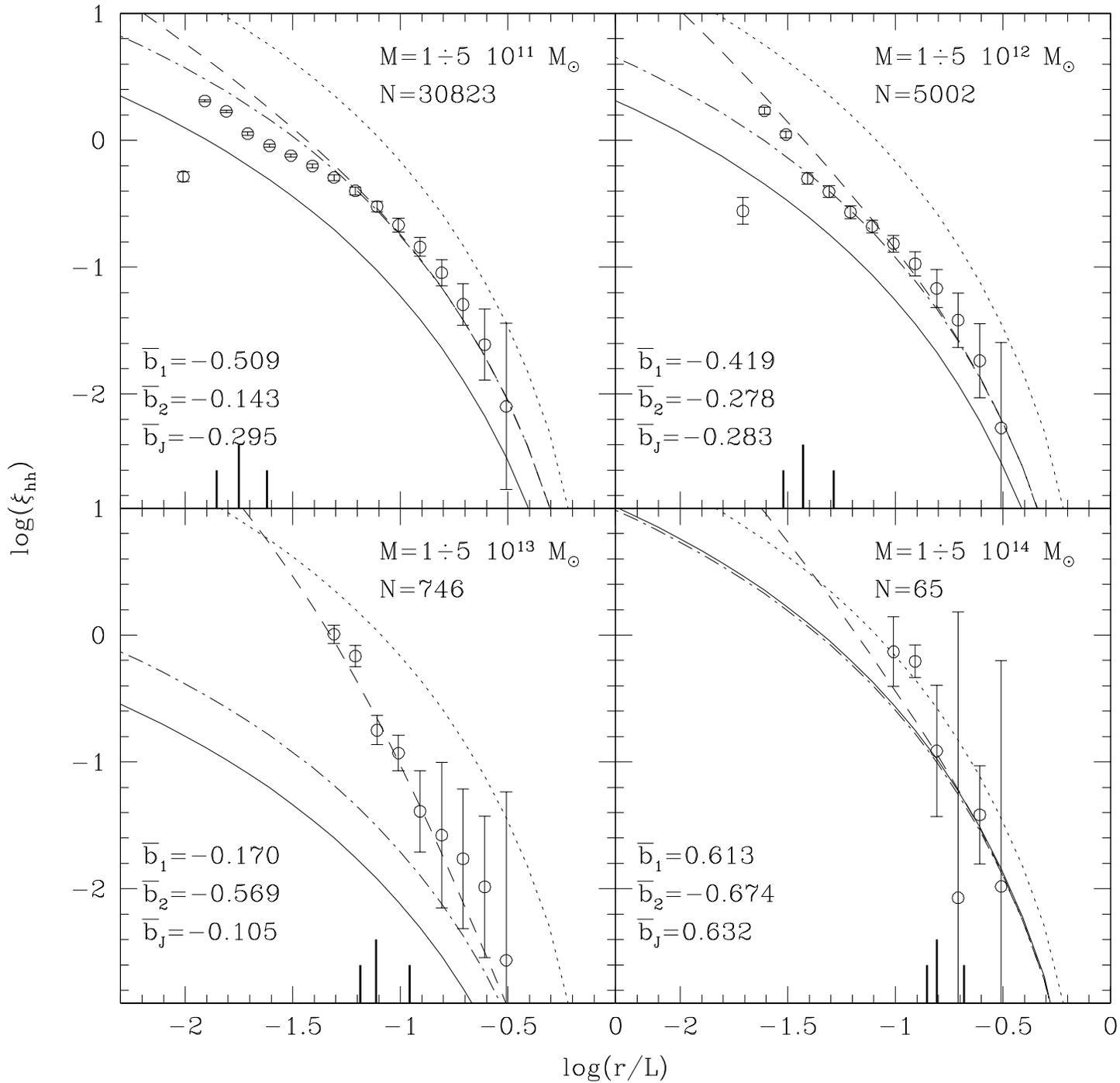
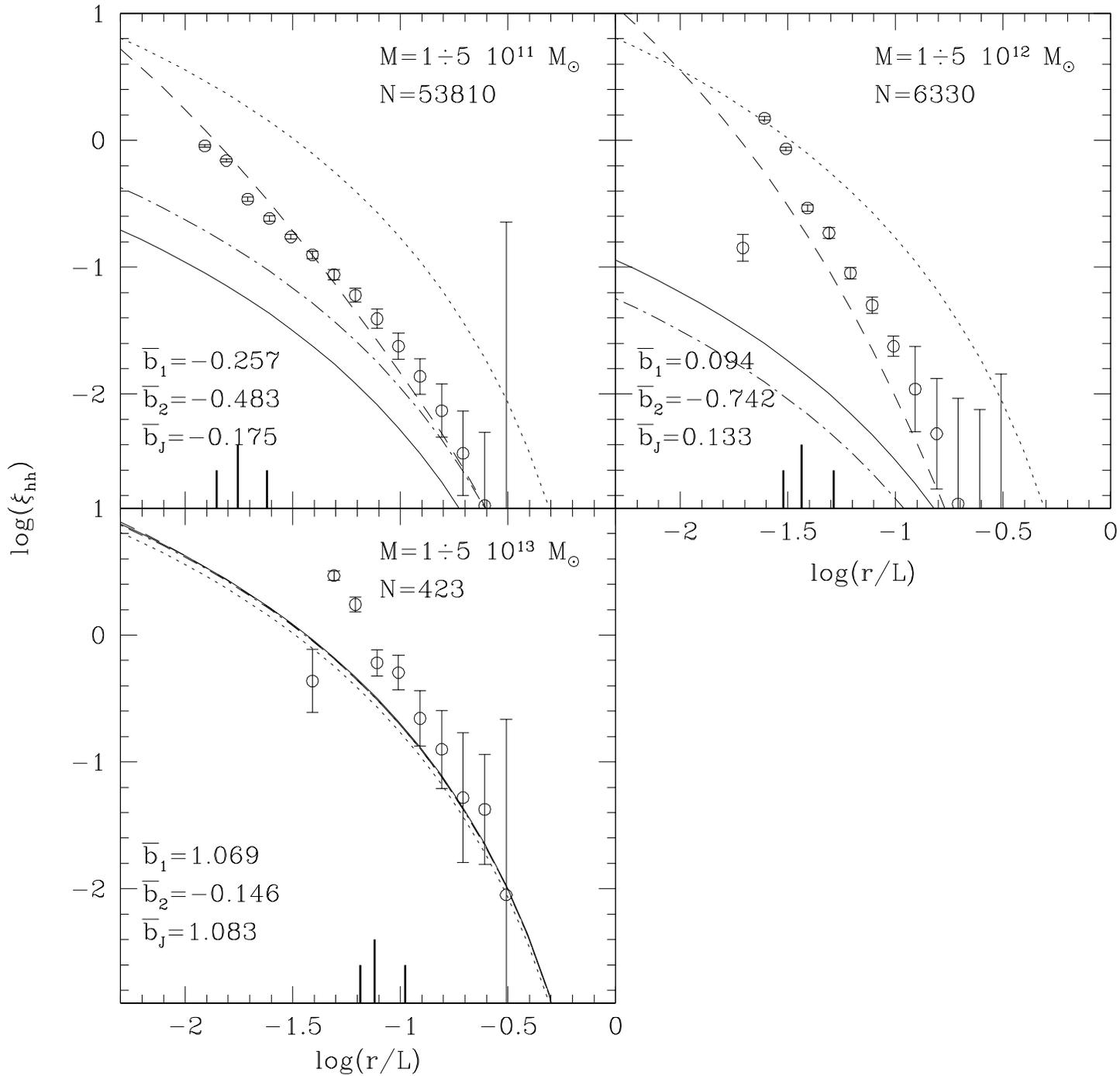
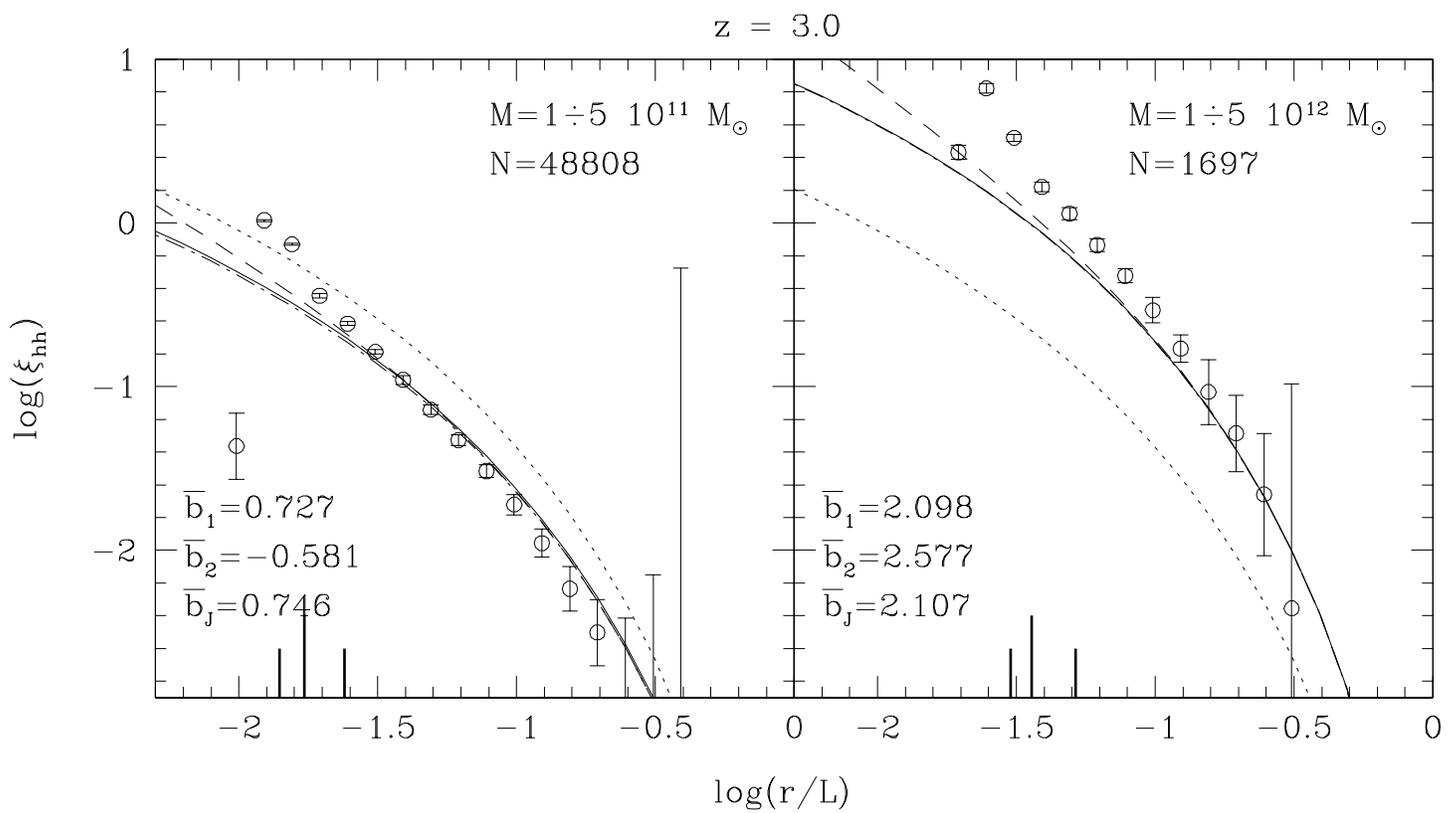
\end{document}